\documentstyle[psfig,graphicx]{mn}

\newif\ifAMStwofonts

\def\simlt{\lower.5ex\hbox{$\; \buildrel < \over \sim \;$}}
\def\simgt{\lower.5ex\hbox{$\; \buildrel > \over \sim \;$}}

\def\apj{ApJ}
\def\apjl{ApJL}
\def\mnras{MNRAS}
\def\pasj{PASJ}
\def\aap{AAp}
\def\aj{AJ}
\def\araa{ARAA}

\ifoldfss
  \ifCUPmtlplainloaded \else
    \NewTextAlphabet{textbfit} {cmbxti10} {}
    \NewTextAlphabet{textbfss} {cmssbx10} {}
    \NewMathAlphabet{mathbfit} {cmbxti10} {} 
    \NewMathAlphabet{mathbfss} {cmssbx10} {} 
  \fi
  \ifAMStwofonts
    \ifCUPmtlplainloaded \else
      \NewSymbolFont{upmath} {eurm10}
      \NewSymbolFont{AMSa} {msam10}
      \NewMathSymbol{\upi}     {0}{upmath}{19}
      \NewMathSymbol{\umu}     {0}{upmath}{16}
      \NewMathSymbol{\upartial}{0}{upmath}{40}
      \NewMathSymbol{\leqslant}{3}{AMSa}{36}
      \NewMathSymbol{\geqslant}{3}{AMSa}{3E}

    \fi
  \fi
\fi 

\ifnfssone
  \newmathalphabet{\mathit}
  \addtoversion{normal}{\mathit}{cmr}{m}{it}
  \addtoversion{bold}{\mathit}{cmr}{bx}{it}
  \newmathalphabet{\mathbfit} 
  \addtoversion{normal}{\mathbfit}{cmr}{bx}{it}
  \addtoversion{bold}{\mathbfit}{cmr}{bx}{it}
  \newmathalphabet{\mathbfss} 
  \addtoversion{normal}{\mathbfss}{cmss}{bx}{n}
  \addtoversion{bold}{\mathbfss}{cmss}{bx}{n}
  \ifAMStwofonts
    \ifCUPmtlplainloaded \else
      \UseAMStwoboldmath
      \makeatletter
      \new@mathgroup\upmath@group
      \define@mathgroup\mv@normal\upmath@group{eur}{m}{n}
      \define@mathgroup\mv@bold\upmath@group{eur}{b}{n}
      \edef\UPM{\hexnumber\upmath@group}
      \new@mathgroup\amsa@group
      \define@mathgroup\mv@normal\amsa@group{msa}{m}{n}
      \define@mathgroup\mv@bold\amsa@group{msa}{m}{n}
      \edef\AMSa{\hexnumber\amsa@group}
      \makeatother
      \mathchardef\upi="0\UPM19
      \mathchardef\umu="0\UPM16
      \mathchardef\upartial="0\UPM40
      \mathchardef\leqslant="3\AMSa36
      \mathchardef\geqslant="3\AMSa3E
    \fi
  \fi
\fi 

\ifnfsstwo
  \DeclareMathAlphabet{\mathbfit}{OT1}{cmr}{bx}{it}
  \SetMathAlphabet\mathbfit{bold}{OT1}{cmr}{bx}{it}
  \DeclareMathAlphabet{\mathbfss}{OT1}{cmss}{bx}{n}
  \SetMathAlphabet\mathbfss{bold}{OT1}{cmss}{bx}{n}
  \ifAMStwofonts
    \ifCUPmtlplainloaded \else
      \DeclareSymbolFont{UPM}{U}{eur}{m}{n}
      \SetSymbolFont{UPM}{bold}{U}{eur}{b}{n}
      \DeclareSymbolFont{AMSa}{U}{msa}{m}{n}
      \DeclareMathSymbol{\upi}{0}{UPM}{"19}
      \DeclareMathSymbol{\umu}{0}{UPM}{"16}
      \DeclareMathSymbol{\upartial}{0}{UPM}{"40}
      \DeclareMathSymbol{\leqslant}{3}{AMSa}{"36}
      \DeclareMathSymbol{\geqslant}{3}{AMSa}{"3E}
    \fi
  \fi
\fi 

\ifCUPmtlplainloaded \else
  \ifAMStwofonts \else 
    \def\upi{\pi}
    \def\umu{\mu}
    \def\upartial{\partial}
  \fi
\fi

\title[The SZ Effect Closer to Home]
    {Observing the Sunyaev-Zel'dovich Effect Closer to Home}
\author[Taylor, Moodley \& Diego]
{James E. Taylor$^{1}$\thanks{email: {\tt jet@astro.ox.ac.uk}}, Kavilan Moodley$^{1,2}$ \& J.M.\ Diego$^{1}$\\
$^{1}$Denys Wilkinson Building, 1 Keble Road, Oxford OX1 3RH, United Kingdom \\
$^{2}$School of Mathematical \& Statistical Sciences,
University of Natal, Durban, 4041, South Africa \\}
\date{\today}
\pubyear{2002}

\begin{document}

\maketitle

\begin{abstract}
Hot gas trapped in a dark matter halo will produce a decrement in the 
surface brightness of the microwave background, the Sunyaev-Zel'dovich (SZ) 
effect. While massive clusters produce the strongest central SZ decrements, 
we point out that a local galaxy halo, specifically the halo of M31, may be 
one of the brightest {\it integrated} SZ sources in the sky. For various 
realistic gas distributions consistent with current X-ray limits, we show 
that the integrated SZ decrement from M31 will be comparable to decrements 
already detected in more distant sources, provided its halo contains an
appreciable quantity of hot gas. A measurement of this decrement would 
provide direct information on the mass, spatial distribution and thermodynamic 
state of hot gas in a low-mass halo, and could place important constraints on 
current models of galaxy formation. Detecting such an extended 
($\sim$ 10\degr), low-amplitude signal will be challenging, but should be 
possible with all-sky SZ maps from satellite missions such as the 
Wilkinson Microwave Anisotropy Probe or the Planck Surveyor. 
\end{abstract}


\section{Introduction}\label{sec:intro}

The mean baryonic density of the Universe is now fairly well determined from 
nucleosynthesis (e.g.\ Burles, Nollett, \& Turner 2001), 
the temperature fluctuation power 
spectrum of the cosmic microwave background (CMB) (Spergel et al.\ 2003), 
models of the Lyman alpha forest at high redshift (e.g.\ Hui et al.\ 2002), 
and the hot gas fraction in clusters (e.g.\ Roussel, Sadat \& Blanchard 2000), 
together with constraints on the total matter density and the Hubble
parameter, to be $\Omega_{\rm b} \simeq 0.044$. 
As discussed by Silk (2003), there is a net shortfall in the baryons
observed at the present day: about 20\% are visible in the form of stars 
and cluster gas (Fukugita, Hogan and Peebles 1998), 20 \% are in the
low-redshift Lyman-$\alpha$ forest (Penton, Shull, \& Stocke 2000), and 
about half of the remainder are predicted by simulations of structure 
formation to be in a diffuse, warm intergalactic medium 
(Cen \& Ostriker 1999; Dav{\' e} et al.\ 2001), for which there is also 
increasing observational evidence (e.g. Mathur, Weinberg, \& Chen 2003). 
The missing 30\% of the baryons, by implication, must be in some other phase, 
at densities and temperatures that are very hard to detect.
Several pieces of observational evidence, including the break in the
X-ray luminosity-temperature ($L_{\rm x}$--$T_{\rm x}$) relation
on the scale of groups (e.g.\ Helsdon \& Ponman 2000; Xue \& Wu 2000) 
and spectroscopic signatures of strong winds in damped Lyman-break systems 
(Shapley et al.\ 2002) suggest that this baryon deficit
is connected with galaxy formation, which may eject a substantial fraction
of the baryons out of dark matter haloes and into the IGM. If it is indeed 
true that galaxies process and eject a large fraction of the baryons in their 
haloes, then this is an important feature of galaxy formation that should be 
confirmed by direct observations of the residual hot gas in galaxy haloes.

In principle, the gas content of a dark matter halo can be determined directly 
from its X-ray emission, but whereas galaxy clusters emit strongly in the 
X-ray, due to their large gas content and high virial temperature, typical 
galaxy haloes will be much fainter X-ray emitters. Furthermore, galaxies 
themselves contain bright X-ray point sources, making emission from an 
extended, gaseous component around them harder to detect. Thus, while the 
non-detection of X-ray emission from a diffuse component in low-mass systems 
has set limits on the mass and mean density of gas surrounding typical 
galaxies (Benson et al.\ 2000), the constraints are relatively weak, and 
large quantities of low-density gas could easily be hiding in low-mass haloes.

The Sunyaev-Zel'dovich (SZ) effect, in which CMB photons are scattered
to higher energies by collisions with hot electrons, provides a more 
sensitive way of detecting gas at low densities. This effect has now been
been measured with high signal-to-noise in many massive clusters 
(e.g.\ Mason, Myers, \& Readhead 2001; Grego et al.\ 2001; 
De Petris et al.\ 2002; see Rees et al.\ 2002 for earlier references)
, and increasing quantities 
of SZ data are becoming available from CMB experiments and dedicated 
instruments (see Carlstrom, Holder \& Reese 2002 for a recent review).
The central decrement produced by the SZ effect is independent of redshift, 
and varies in amplitude roughly as $M$, the mass of the cluster,
so current SZ surveys generally focus on 
detecting the most massive galaxy clusters, irrespective of distance. 
The total SZ flux scattered by
a source scales as $M^{5/3}\,d_{a}^{-2}$, however, where $d_{a}$ is the 
angular diameter distance to the source. For a sufficiently 
nearby source, the larger apparent diameter can compensate for the mass 
dependence, making the gas in low-mass systems easier to detect.

In this letter, we suggest that one of the strongest 
integrated SZ signals may come from a galaxy halo, specifically the halo
of our immediate neighbour M31. If this signal could be detected, it
would offer a unique opportunity to study halo gas on a mass-scale that 
is very difficult to probe by other means. In what follows, 
we will first make a 
simple estimate of how the integrated (thermal) SZ flux and X-ray luminosity 
scale 
with mass and distance to the source, and compare the total signal from 
various nearby haloes. We will then calculate more precise values for M31, 
using a realistic model for the halo gas distribution. 
Finally, we will discuss the possibility of detecting a very
low-amplitude temperature decrement from such an extended source, 
using data from all-sky CMB missions such as the
Wilkinson Microwave Anisotropy Probe (WMAP) or the Planck Surveyor. 

\section{The SZ flux from Nearby Haloes}

\subsection{Basic Scaling}

The thermal SZ effect is the temperature decrement produced when CMB photons 
are inverse-Compton scattered off hot gas along the line of sight. The 
relative decrement is proportional to the integral of the gas pressure, 
$P_{\rm e} = n_{\rm e} T_{\rm e}$, along the line of sight. 
The integrated SZ flux, 
that is the total flux scattered in our direction by gas in a particular halo, 
is:  
\begin{equation}
S(\nu)_{\rm SZ} \propto g(\nu) I_o(\nu) \int d\Omega \int_{0}^{r_{\rm vir}} n_{\rm e}({\bf r}) T_{\rm e}({\bf r}) dl 
\end{equation}\label{szflux}
where the function $g(\nu)$ describes the frequency dependence of the SZ 
effect, $I_o(\nu)$ 
is the intensity of the CMB, $d\Omega$ is the solid angle subtended by 
the gas, $r_{\rm vir}$ is the virial radius of the halo,
$\theta$ is the angle between the halo centre and the line of sight,
and $l = (\left|r\right|^2 - \theta^2 d_{a}^2)^{1/2}$.

We can get a rough estimate of the magnitude of the SZ flux
by assuming that the gas is isothermal, and that its temperature
$T_{\rm e}$ is proportional to the virial temperature,
$T_{\rm vir} \propto M^{2/3}$.
Integrating out to the virial radius $r_{\rm vir} \propto M^{1/3}$, 
we then obtain the scaling relation $S(\nu)_{\rm SZ} \propto 
M^{5/3} d_{a}^{-2}(z)$, where $d_{a}(z)$ is the (angular diameter) 
distance to the source. 

\begin{figure}
\includegraphics[width=84mm]{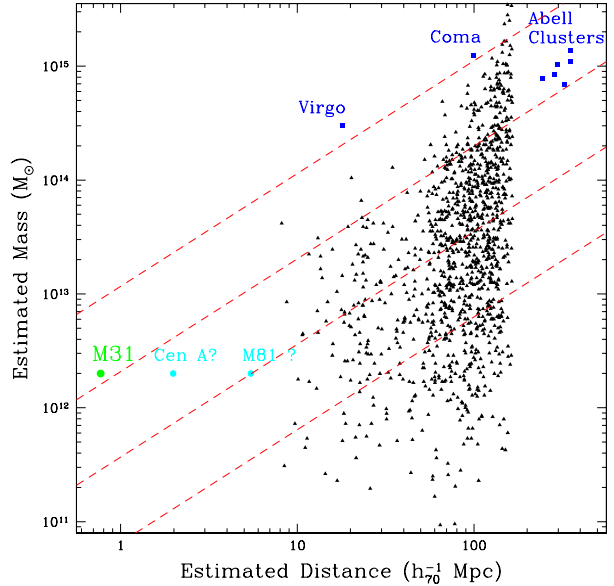}
  \caption[]{Estimated (angular diameter) distances and masses for 
nearby galaxy, group and cluster haloes. 
The small triangles are from the UZC-SSRS2 catalogue (Ramella et al.\ 2002)
of nearby groups and clusters, 
while the large points show the loci of specific 
objects with distances and halo masses estimated by various means 
(see text for details). Lines indicate constant integrated SZ flux, 
assuming a simple scaling with mass and distance.}
\label{fig1}
\end{figure}

Fig.\ \ref{fig1} shows the estimated mass of various haloes, versus their 
estimated distance from us. The small black triangles are nearby groups or 
clusters from the UZC-SSRS2 catalogue (Ramella et al.\ 2002), with estimated 
distances derived from the redshift assuming zero peculiar velocity, and 
estimated masses as given in the catalogue. We also include various nearby 
galaxy and cluster haloes (large points), with their names indicated. The 
distances and masses for these objects were taken from the 
Lyon-Meudon Extragalactic Database (LEDA) and from various recent references 
(Virgo -- Schindler, Binggeli, \& B{\" o}hringer 1999; 
Fouqu{\' e} et al.\ 2001; Coma -- Liu \& Graham 2001). For M81 and Cen A, 
we assumed the same halo mass as M31. The more distant points are a 
set of Abell clusters for which the SZ effect has been measured 
(Mason et al.\ 2001), with distances based on their redshift and masses 
scaled up from the mass of Coma by $(T_{\rm Abell}/T_{\rm Coma})^{3/2}$. 
We assume a Hubble constant of 70\,km\,s$^{-1}$\,Mpc$^{-1}$ in all cases.

Both the mass and the distance estimates plotted here contain substantial 
uncertainties. Nonetheless, they give a general indication of representative 
values for nearby haloes. Also indicated on the plot are lines of constant 
integrated SZ flux, assuming the scaling 
$S(\nu)_{\rm SZ} \propto M^{5/3} d_{a}^{-2}$.
We see that Coma stands out as one of the most massive nearby clusters;
indeed, it was recently detected at about the 10-$\sigma$ level 
(that is 5, 0.5 and 5-$\sigma$ in 3 separate bands) 
by De Petris et al.\ (2002), for instance. We also see that 
there are various other promising targets for SZ observations, 
including Virgo and the Abell clusters. 
At very small distances, however, another source stands out: the halo of M31. 
Although a thousand times less massive than the Abell clusters, it is close 
enough that it may produce a comparable SZ flux locally.

An extended halo around M31 might also be detectable in X-rays.
We can estimate the X-ray luminosity from gas in the halo of M31, for the
simple isothermal gas distribution considered above.
The total X-ray luminosity is:
\begin{equation}
L_{\rm X} = \int dV\,n_{\rm e}^2\Lambda(T) dr\,.
\end{equation}\label{fluxm0}
Here $\Lambda(T)$ is the X-ray emissivity, which is consists
of a bremsstrahlung term scaling as $T^{1/2}$, and a term due to spectral
lines. For simplicity, we will assume the line emission has the same 
scaling with $T$, and increases the total luminosity by a factor of roughly 
2 (as appropriate for a low metallicity gas in the 0.5-2.0 keV band). 
The total luminosity should therefore scale as 
$L_x \propto M n_{\rm e}(0)^2 T^{1/2} \propto M^{4/3} n_{\rm e}(0)^2$. 
Thus, even a massive halo of gas may escape detection provided 
it is sufficiently extended, so that $n_{\rm e}$ is low. 

In fact, recent ASCA observations by Takahashi et al.\ (2001) 
place an upper limit of $2.3 \pm 0.3 \times 10^{38}$ erg s$^{-1}$
on soft (0.5-10 keV) emission from within 12' 
(or 2.7\,kpc at a distance of 770\,kpc) of the centre of M31,
while an earlier analysis of ROSAT data by 
West, Barber and Folgheraiter (1997)
limited the contribution to the luminosity from within 
21.5\,kpc (at 770\,kpc) to less than
$5.4 \times 10^{39}$ erg s$^{-1}$ in the 0.5--2.0 keV band, 
assuming M31 has the same mass as the Milky Way. We will
scale this mass up to $10^{40}$ erg s$^{-1}$,
on the assumption that M31 is at least 50 per cent 
more massive than the Milky Way, 
as implied by its greater circular velocity.
As we will show below, these limits do not constrain the
mass of gas in the halo of M31 very strongly, provided the gas
is distributed well beyond the optical extent of the galaxy.

So far in this discussion we have assumed simple scalings
for the sizes, temperatures and densities of halo gas distributions.
Observations show that real systems deviate from these simple
assumptions. The relationship between X-ray luminosity and temperature
is a power-law for massive clusters, for instance, but changes
slope on group and galaxy scales (e.g.\ Helsdon \& Ponman 2000; 
Xue \& Wu 2000 and references therein). Overall, low-mass haloes 
are less X-ray luminous than expected from simple scalings or from
galaxy formation models (Benson et al.\ 2000). This could indicate 
that low-mass
haloes contain little hot gas, their primordial gas having cooled 
and/or having been ejected at some point in the halo's history. In this case, 
the SZ effect may be undetectable on galaxy scales. Alternatively, however,
the gas may be present in an extended, low-density distribution. In this 
case, the X-ray signal will be reduced, but the total SZ flux will be 
unaffected. We investigate these possibilities in more detail below.

\subsection{Simple Halo Models}

We will now consider more detailed models
of the gas distribution in the halo of M31, to calculate 
realistic values for the total SZ decrement and X-ray luminosity.
The distribution of hot gas in galaxy clusters is reasonably well
described by a empirically-derived fitting function, the beta profile, 
\begin{equation}
n_{\rm e}(r) = n_{\rm e}(0)(1 + (r/r_{\rm c})^2)^{-3\beta/2}
\end{equation}
(Cavaliere \& Fusco-Femiano 1978).
We will assume $\beta = 2/3$, since this is the value which best matches
cluster X-ray profiles (Jones \& Foreman 1984; 
Mohr, Mathiesen, \& Evrard 1999).
For this particular model, if the gas is isothermal, then the 
integrated SZ flux from within a projected radius $R$ is:
\vspace{10mm}
\begin{eqnarray}
S_{\rm SZ}(R,\nu) = \hspace{6cm}\nonumber\\
\hspace{6mm}S_0\,g(\nu)\,\left({n_{\rm e}(0)\over{10^{-3}{\rm cm}^{-3}}}\right)\left({T_{\rm e}\over{1\,{\rm keV}}}\right)\left({r_{\rm c}\over{1\,{\rm kpc}}}\right)^3\,f_1(R)\,,
\end{eqnarray}
where $S_0 = 3.615\times 10^{-3}$ Jy 
(assuming a distance of 770\,kpc to M31), 
$g(\nu)$ is a function of order 1 describing
the frequency dependence of the effect,
and $f_1(R) = [(1 + (R/r_{\rm c})^2)^{1/2} - 1]$.
The contribution to the bolometric X-ray luminosity from within the same 
radius is:
\begin{eqnarray}
L_{\rm X}(R) = \hspace{62mm}\nonumber\\
L_{\rm X,0}\left({n_{\rm e}(0)\over{10^{-3}{\rm cm}^{-3}}}\right)^2\left({T\over{1\,{\rm keV}}}\right)^{0.5}\left({r_{\rm c}\over{1\,{\rm kpc}}}\right)^3\,f_2(R)\,,
\end{eqnarray}
where $L_{\rm X,0} = 1.10\times 10^{35}$erg\,s$^{-1}$ (assuming the dominant
contribution is from free-free emission, and that line emission is negligible) 
and $f_2(R) = [1 - (1 + (R/r_{\rm c})^2)^{-1/2}]$.
The electron density is normalised by specifying the mass within some radius
$M_{\rm hot}(<R) = 4\,\pi\, m_{\rm p}\,\mu\,n_{\rm e}(0) r_{\rm c}^{-3} [(R/{r_{\rm c}}) - {\rm tan}^{-1}(R/{r_{\rm c}})]$.
We note that the gas distribution in simulated haloes drops off
faster than $r^{-2}$ in the outer parts, so if we integrated the beta 
profile out to the virial radius of the halo we would overestimate the mass
in the outer regions. To avoid this complication, we consider only the
mass within the central 100\,kpc of the halo, and use this to set $n_{\rm e}(0)$.

To determine the amplitude of the SZ and X-ray emission, we need to fix the 
values of $T$, $M_{\rm hot}(<100\,{\rm kpc})$, and $r_{\rm c}$. 
We will assume that the gas in
the halo of M31 follows the mass-temperature relation observed for slightly
more massive group haloes, so that $T \simeq 0.5$ keV. For 
$M_{\rm hot}(<100\,{\rm kpc})$, the total mass of hot gas within the central
100\,kpc, we have the constraint that the
baryon fraction of the whole halo not exceed the universal value 
$f_{\rm b} = \Omega_{\rm b}/\Omega_{\rm m} = 0.166$ 
(Spergel et al.\ 2003). The mass
of the halo of M31 is estimated to be between 1 and 
$2 \times 10^{12} M_{\odot}$ (Klypin, Zhao, \& Somerville 2002). 
Thus its total, primordial baryonic mass
would have been $1.66-3.32 \times 10^{11} M_{\odot}$, of which approximately
$9 \times 10^{10} M_{\odot}$ are in the form of stars and gas in the galaxy 
itself. This leaves up to $\simeq 2.4 \times 10^{11} M_{\odot}$ 
of hot gas in the halo. If this gas traced the dark matter distribution,
roughly half of it would lie in the central 100 kpc of the system
(assuming a virial radius of 330 kpc). Then again, radiative cooling
or galactic heating may have altered the gas distribution, causing it
to be more or less centrally condensed. We will consider values of 
$M_{\rm hot}$ between $10^{10}$ and $1.5 \times 10^{11}\,M_{\odot}$,
which should cover a realistic range in mass.
 
Finally, we have to choose a value for the core radius of the gas 
distribution, 
$r_{\rm c}$. Detailed models of cluster gas suggest that this should 
be roughly equal to the core radius of the dark matter distribution, 
$r_{\rm c,DM}$\,,
which is 27.5\,kpc for a halo mass of $2 \times 10^{12} M_{\odot}$
in a LCDM cosmology (Eke, Navarro \& Steinmetz 2001). 
On the other hand, there has been recent observational
evidence for a much larger, lower density distribution of
gas around the Milky Way, extending out to 70\,kpc or more 
(Sembach et al.\ 2002). Given this uncertainty, for each model we will 
choose the smallest core radius consistent with the X-ray  
limits discussed previously, which will generally be larger than 
$r_{\rm c,DM}$.
Slightly larger or smaller core radii will have little effect on the
detectability of the SZ signal, which depends mainly on $M_{\rm hot}$,
as shown below.

\begin{center}
\begin{table*}
\begin{minipage}{150mm}
\caption{Model Gas Distributions \label{tbl-1}}
\begin{tabular}{lccccccc}
\hline
Model&$M_{\rm hot}$
&$r_{\rm c}$&$r_{\rm c}$
&$L_{\rm X}(< r_1)$&$L_{\rm X}(< r_2)$
&$S_{\rm SZ}(< r_{\rm c})$&$S_{\rm SZ}(< 100 {\rm\,kpc})$\\
&($10^{11} M_{\odot}$)&(kpc)&($^{\circ}$)
&($10^{38}$ erg\, s$^{-1}$)&($10^{38}$ erg\, s$^{-1}$)
&(Jy @ 150 GHz)&(Jy @ 150 GHz)\\
\hline
A&1.5&57&4.2&1.6&94&-11.3&-27.8 \\
B&1.0&48&3.6&1.7&95&-7.5&-23.8 \\
C&0.5&36&2.7&1.8&91&-3.8&-17.7 \\
D&0.3&29&2.2&1.9&87&-2.3&-14.1 \\
E&0.2&24&1.8&2.2&89&-1.5&-11.9 \\
F&0.1&18&1.3&2.3&74&-0.8&-8.4 \\
\hline
\end{tabular}
\end{minipage}
\end{table*}
\end{center}

In table \ref{tbl-1} we give the SZ and X-ray emission 
from various model haloes, 
with parameters chosen based on these arguments. We calculate the integrated
SZ flux from within the core radius and 100\,kpc, as well as the X-ray luminosity
from the two regions for which there are observational limits, $r_1 = 2.7$\,kpc
and $r_2 = 21.5$\,kpc.
We see that the SZ flux from some models can be very large, while the total
X-ray luminosity is still below the observational limits. To determine 
whether any of these gas distributions would be detected, however, we have 
to estimate the noise level in SZ maps filtered on large scales. 
We will make a rough calculation of the signal-to-noise
in the next section. First, however, we consider the additional contribution 
to the signal from the kinetic SZ effect.

\subsection{Kinetic SZ effect}

Motion of gas with respect to the CMB also produces a second, kinetic SZ effect
with a distinct spectral signature. The amplitude of the kinetic effect 
relative to the thermal effect is:
\begin{equation}
S^{\rm k}_0/S^{\rm t}_0\,(\nu) = \left({1.67}\over{j(\nu)}\right) \left({{\rm v}_{\rm CMB}}\over{1000\,{\rm km\,s}^{-1}}\right) \left({T_{\rm e}}\over{1.0\,{\rm keV}}\right)^{-1}\,,
\end{equation}
where $j(\nu)$ is a function of order one describing
the spectral dependence
and ${\rm v}_{\rm CMB}$ is the velocity of the gas with respect to the CMB.
For clusters, where $T_{\rm e}$ is large, this ratio is typically less than 10 
per cent, but in M31 $T_{\rm e}$ would be much lower, and thus the kinetic
effect could boost the SZ signal substantially. If we assume the gas is 
moving with the same velocity as the central galaxy (whose motion is almost
opposite to the CMB dipole), then ${\rm v}_{\rm CMB} = 582$\,km\,s$^{-1}$, 
and at 150 GHz, $S^{\rm k}_0/S^{\rm t}_0 \simeq 2$, so the combined 
SZ flux will be three times larger at this frequency. 
The halo of M31 may also rotate at 
an appreciable velocity, of course, which would modulate this contribution. 
Finally, we note that the thermal and kinetic contributions to this signal 
can in principle be separated, due to their different frequency dependence.
Thus it could provide the first strong detection of the kinetic SZ effect.
This separation will require good spatial resolution and excellent 
signal-to-noise, however, since the frequency dependence of the kinetic
SZ signal is the same as that of the CMB.

\subsection{Detection Strategies}

\begin{figure}
\includegraphics[width=80mm]{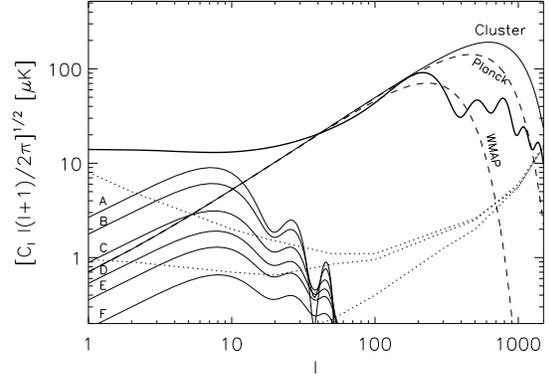}  
  \caption[]{The angular power spectrum (RJ)
    for the models listed in table \ref{tbl-1}
(thin solid curves), compared with the power spectrum of primordial 
anisotropies (thick curve), and various estimates of the 
(RJ) residuals in Planck maps (dotted lines; Tegmark et al.\ 2000).
Also shown is the power spectrum for a typical Abell cluster, and the
same spectrum convolved with the WMAP and Planck beams (dashed curves).}
\label{fig2}
\end{figure}

For the models considered above, we expect a total SZ 
flux of $\simeq$ 1--10 Janskys or more, spread out over a patch several 
degrees in size. While this signal is comparable to those detected from 
much more compact sources such as distant clusters, it may be harder to 
extract from SZ maps contaminated by residuals from primary anisotropies 
or large-scale foregrounds. We have estimated the contribution of our models 
to the power spectrum of the Rayleigh-Jeans (RJ) regime in the Planck maps. 
Since the power spectrum is an 
average quantity over the sky, our estimates need to be re-scaled by an area 
correction factor in order to make a fair comparison. In Fig.\ \ref{fig2},
we compare the contribution from the models to estimates of the residuals 
in Planck SZ maps (Tegmark et al.\ 2000), making pessimistic, realistic, or 
optimistic assumptions about the efficiency of foreground subtraction 
(dotted lines, from top to bottom). We see that for realistic assumptions
about foreground subtraction (middle dotted line), our models are detectable
at the 1--10$\sigma$ level on 20$^{\circ}$ scales. The kinetic contribution
could increase the signal by an additional factor of 2--3, provided
it could be separated from the CMB itself.

\begin{figure}
\centerline{\includegraphics[width=50mm]{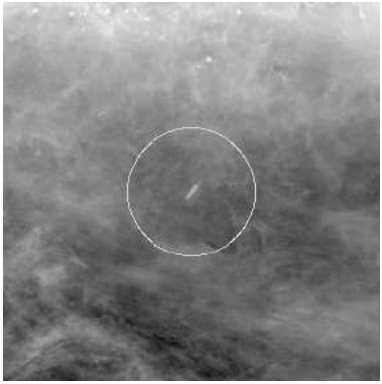}}  
\caption[]{A section of the SFD dust map (Schlegel et al.\ 1998), centred on M31.
The patch within 100\,kpc of the galaxy (indicated by the circle)
is relatively free of foreground dust emission. }
\label{fig3}
\end{figure}

We note that M31 lies in a zone of reasonably low foregrounds. 
Fig.\ \ref{fig3} shows a section of the SFD dust map 
(Schlegel, Finkbeiner, \& Davis 1998) centred on M31. The circle 
indicates a projected radius of 7.4$^{\circ}$, or 100\,kpc at a
distance of 770\,kpc. Since we know the centre
of the halo gas distribution a priori, we may be able to optimise
the SZ detection further, by using filters centred on this point
or focusing on the parts of this region less contaminated by foregrounds. 
Given the low amplitude and large spatial extent of the source, the 
signal from M31 will probably not be detectable by interferometers.
The data from the WMAP mission is more promising in this regard,
although it is also at lower frequency where the relative 
contribution from the thermal SZ effect is smaller.
Recent analysis by Tegmark, de Oliveira-Costa \& Hamilton (2003)
finds that WMAP residuals have an r.m.s.\ amplitude of 4--5 $\mu K$ 
at $l = 10$ in the cleanest regions of the sky. 
Using a naive combination of WMAP frequencies (Q-W bands) to subtract 
the CMB, we expect roughly 6 times less signal from the SZ effect 
than in the RJ regime assumed in Fig.\ \ref{fig2}, 
so even model `A' would not be detectable above this noise level
without using more elaborate methods. On the other hand, the 4-year
WMAP data may start to rule out the most massive of our halo models.

\section{Discussion}

We have shown that for a series of realistic models, the halo of
M31 should produce a substantial integrated SZ flux. While the
precise extent and amplitude of the flux depend on the details
of the gas distribution, more elaborate models should produce roughly
the same values as calculated here. In particular, since the SZ effect 
depends on the integral of the electron pressure along the line of sight,
we expect the same total flux even if the pressure support for the halo
gas close to M31 itself comes from a hotter, lower-density component
heated by galactic sources, as suggested by X-ray observations 
(e.g.\ Takahashi et al.\ 2001).

From the calculations above, the integrated SZ flux should be detectable at 
the 1-$\sigma$ to 10-$\sigma$ level in SZ maps based on Planck data, provided 
that the foreground removal is reasonably accurate, and that the halo of
M31 still contains a reasonable amount of hot gas. If this signal is detected, 
it will provide a direct
determination of the thermodynamic properties of halo gas on galaxy scales. 
Conversely, if the SZ effect is not detected it will place useful constraints 
on the amount of gas left over after the formation of a spiral galaxy within
this halo. We do not expect the SZ signal from M31 to be visible in the 
first-year WMAP data, but the 4-year data should also start to constrain the 
gas distribution in its halo. 
Thus CMB experiments may soon provide an unexpected view of 
the effects of galaxy formation close at hand. 

\vspace{-0.6cm}

\section*{Acknowledgements}
The authors would like to thank Nabila Aghanim, Greg Bryan, 
Marc Sarzi and Joe Silk for useful discussions. JET acknowledges support from 
the Leverhulme Trust during the course of this work. KM acknowledges the
support of a PPARC Fellowship (PPA/G/S/1999/00671). JMD acknowledges the 
support of a Marie Curie Fellowship from the European Community 
(HPMF-CT-200-00967).
The dust map shown in Fig.\ \ref{fig3} was generated using NASA's
SkyView facility, located at NASA Goddard Space Flight Center 
(http://skyview.gsfc.nasa.gov).


\vspace{-0.6cm}

\begin{thebibliography}{}

\bibitem[Benson, Bower, Frenk, \& White(2000)]{2000MNRAS.314..557B} Benson, 
A.~J., Bower, R.~G., Frenk, C.~S., \& White, S.~D.~M.\ 2000, \mnras, 314, 
557 

\bibitem[Burles, Nollett, \& Turner(2001)]{2001ApJ...552L...1B} Burles, S., 
Nollett, K.~M., \& Turner, M.~S.\ 2001, \apjl, 552, L1 

\bibitem[Carlstrom, Holder, \& Reese(2002)]{2002ARA&A..40..643C} Carlstrom, 
J.~E., Holder, G.~P., \& Reese, E.~D.\ 2002, \araa, 40, 643 

\bibitem[Cavaliere \& Fusco-Femiano(1978)]{1978A&A....70..677C} Cavaliere, 
A.~\& Fusco-Femiano, R.\ 1978, \aap, 70, 677 

\bibitem[Cen \& Ostriker(1999)]{1999ApJ...514....1C} Cen, R.~\& Ostriker, 
J.~P.\ 1999, \apj, 514, 1 

\bibitem[Dav{\' e} et al.(2001)]{2001ApJ...552..473D} Dav{\' e}, R.~et al.\ 
2001, \apj, 552, 473 

\bibitem[De Petris et al.(2002)]{2002ApJ...574L.119D} De Petris, M.~et al.\ 
2002, \apjl, 574, L119 

\bibitem[Eke, Navarro, \& Steinmetz(2001)]{2001ApJ...554..114E} Eke, V.~R., 
Navarro, J.~F., \& Steinmetz, M.\ 2001, \apj, 554, 114 

\bibitem[Fouqu{\' e}, Solanes, Sanchis, \& 
Balkowski(2001)]{2001A&A...375..770F} Fouqu{\' e}, P., Solanes, J.~M., 
Sanchis, T., \& Balkowski, C.\ 2001, \aap, 375, 770 

\bibitem[Fukugita, Hogan, \& Peebles(1998)]{1998ApJ...503..518F} Fukugita, 
M., Hogan, C.~J., \& Peebles, P.~J.~E.\ 1998, \apj, 503, 518 

\bibitem[Grego et al.(2001)]{2001ApJ...552....2G} Grego, L., Carlstrom, 
J.~E., Reese, E.~D., Holder, G.~P., Holzapfel, W.~L., Joy, M.~K., Mohr, 
J.~J., \& Patel, S.\ 2001, \apj, 552, 2 

\bibitem[Helsdon \& Ponman(2000)]{2000MNRAS.315..356H} Helsdon, S.~F.~\& 
Ponman, T.~J.\ 2000, \mnras, 315, 356 

\bibitem[Hui, Haiman, Zaldarriaga, \& Alexander(2002)]{2002ApJ...564..525H} 
Hui, L., Haiman, Z., Zaldarriaga, M., \& Alexander, T.\ 2002, \apj, 564, 
525 

\bibitem[Jones \& Forman(1984)]{1984ApJ...276...38J} Jones, C.~\& Forman, 
W.\ 1984, \apj, 276, 38 

\bibitem[Klypin, Zhao, \& Somerville(2002)]{2002ApJ...573..597K} Klypin, 
A., Zhao, H., \& Somerville, R.~S.\ 2002, \apj, 573, 597 

\bibitem[Liu \& Graham(2001)]{2001ApJ...557L..31L} Liu, M.~C.~\& Graham, 
J.~R.\ 2001, \apjl, 557, L31 

\bibitem[Mason, Myers, \& Readhead(2001)]{2001ApJ...555L..11M} Mason, 
B.~S., Myers, S.~T., \& Readhead, A.~C.~S.\ 2001, \apjl, 555, L11 

\bibitem[Mathur, Weinberg, \& Chen(2003)]{2003ApJ...582...82M} Mathur, S., 
Weinberg, D.~H., \& Chen, X.\ 2003, \apj, 582, 82 

\bibitem[Mohr, Mathiesen, \& Evrard(1999)]{1999ApJ...517..627M} Mohr, 
J.~J., Mathiesen, B., \& Evrard, A.~E.\ 1999, \apj, 517, 627 

\bibitem[Penton, Shull, \& Stocke(2000)]{2000ApJ...544..150P} Penton, 
S.~V., Shull, J.~M., \& Stocke, J.~T.\ 2000, \apj, 544, 150 

\bibitem[Ramella, Geller, Pisani, \& da Costa(2002)]{2002AJ....123.2976R} 
Ramella, M., Geller, M.~J., Pisani, A., \& da Costa, L.~N.\ 2002, \aj, 123, 
2976 

\bibitem[Reese et al.(2002)]{2002ApJ...581...53R} Reese, E.~D., Carlstrom, 
J.~E., Joy, M., Mohr, J.~J., Grego, L., \& Holzapfel, W.~L.\ 2002, \apj, 
581, 53 

\bibitem[Roussel, Sadat, \& Blanchard(2000)]{2000A&A...361..429R} Roussel, 
H., Sadat, R., \& Blanchard, A.\ 2000, \aap, 361, 429 

\bibitem[Schindler, Binggeli, \& B{\" o}hringer(1999)]{1999A&A...343..420S} 
Schindler, S., Binggeli, B., \& B{\" o}hringer, H.\ 1999, \aap, 343, 420 

\bibitem[Schlegel, Finkbeiner, \& Davis(1998)]{1998ApJ...500..525S} 
Schlegel, D.~J., Finkbeiner, D.~P., \& Davis, M.\ 1998, \apj, 500, 525 

\bibitem[Sembach et al.(2002)]{mwhotgas} Sembach, K.~R., et al.\ 2002, 
ApJ, submitted (astro-ph/0207562)

\bibitem[Shapley et al.(2000)]{lyalpha}Shapley, A. Steidel, C. Adelberger, K.
\& Pettini, M. 2002, in `A New Era in Cosmology', T. Shanks \& N. Metcalf eds.,
in press (astro-ph/0107324)

\bibitem[Silk(2003)]{silk2003} Silk, 
J., 2003, \mnras, submitted (astro-ph/0212068) 

\bibitem[Spergel et al.(2003)]{WMAP} Spergel, D.~N., et al.\ 2003, 
ApJ submitted (astro-ph/0302209)

\bibitem[Takahashi, Okada, Kokubun, \& 
Makishima(2001)]{2001PASJ...53..875T} Takahashi, H., Okada, Y., Kokubun, 
M., \& Makishima, K.\ 2001, \pasj, 53, 875 

\bibitem[Tegmark, Eisenstein, Hu, \& de 
Oliveira-Costa(2000)]{2000ApJ...530..133T} Tegmark, M., Eisenstein, D.~J., 
Hu, W., \& de Oliveira-Costa, A.\ 2000, \apj, 530, 133 

\bibitem[Tegmark, de Oliveira-Costa \& Hamilton (2003)]{teg2} Tegmark, M., 
de Oliveira-Costa, A., \& Hamilton, A.~J.~S.\ 2003, PRD, submitted
(astro-ph/0302496)

\bibitem[West, Barber, \& Folgheraiter(1997)]{1997MNRAS.287...10W} West, 
R.~G., Barber, C.~R., \& Folgheraiter, E.~L.\ 1997, \mnras, 287, 10 

\bibitem[Xue \& Wu(2000)]{2000ApJ...538...65X} Xue, Y.~\& Wu, X.\ 2000, 
\apj, 538, 65 

\end{thebibliography}
\end{document}